\documentstyle[12pt]{article}

\newcommand{\zed}{{\bb Z}} 
\newcommand{\real}{{\bb R}} 
\newcommand{\id}{{\bb I}} 
\newcommand{\alg}{{\cal A}} 

\font\mybb=msbm10 at 12pt
\def\bb#1{\hbox{\mybb#1}}

\def\e{{\rm e}}

\def\beq{\begin{equation}}
\def\eeq{\end{equation}}
\def\bea{\begin{eqnarray}}
\def\eea{\end{eqnarray}}
\def\bd{\begin{displaymath}}
\def\ed{\end{displaymath}}

\makeatletter
\newdimen\normalarrayskip              
\newdimen\minarrayskip                 
\normalarrayskip\baselineskip
\minarrayskip\jot
\newif\ifold             \oldtrue            
\def\arraymode{\ifold\relax\else\displaystyle\fi} 
\def\@arrayskip{\ifold\baselineskip\z@\lineskip\z@
     \else
     \baselineskip\minarrayskip\lineskip2\minarrayskip\fi}
\def\@arrayclassz{\ifcase \@lastchclass \@acolampacol \or
\@ampacol \or \or \or \@addamp \or
   \@acolampacol \or \@firstampfalse \@acol \fi
\edef\@preamble{\@preamble
  \ifcase \@chnum
     \hfil$\relax\arraymode\@sharp$\hfil
     \or $\relax\arraymode\@sharp$\hfil
     \or \hfil$\relax\arraymode\@sharp$\fi}}
\def\@array[#1]#2{\setbox\@arstrutbox=\hbox{\vrule
     height\arraystretch \ht\strutbox
     depth\arraystretch \dp\strutbox
     width\z@}\@mkpream{#2}\edef\@preamble{\halign \noexpand\@halignto
\bgroup \tabskip\z@ \@arstrut \@preamble \tabskip\z@ \cr}%
\let\@startpbox\@@startpbox \let\@endpbox\@@endpbox
  \if #1t\vtop \else \if#1b\vbox \else \vcenter \fi\fi
  \bgroup \let\par\relax
  \let\@sharp##\let\protect\relax
  \@arrayskip\@preamble}
\makeatother

\newlength{\extraspace}
\newlength{\extraspaces}
\begin{document}

\renewcommand{\footnotesize}{\small}

\addtolength{\baselineskip}{.8mm}

\thispagestyle{empty}

\begin{flushright}
{\sc OUTP}-97-23P\\
DSF/24-97\\
hep-th/9706107\\
\hfill{  }\\ \today
\end{flushright}
\vspace{.3cm}

\begin{center}
{\large\sc{Target Space Duality in Noncommutative Geometry}}\\[15mm]

{\sc Fedele Lizzi}\footnote{Permanent Address: Dipartimento di Scienze Fisiche,
Universit\`a di Napoli Federico II and INFN, Sezione di Napoli, Italy.} {\sc
and Richard J.\ Szabo}\footnote{Work supported in part by the Natural Sciences
and Engineering Research Council of Canada.} \\[2mm]
{\it Department of Physics -- Theoretical Physics\\ University of Oxford\\ 1
Keble Road, Oxford OX1 3NP, U.K.} \\[15mm]

\vskip 0.5 in

{\sc Abstract}

\begin{center}
\begin{minipage}{14cm}

The structure of spacetime duality and discrete worldsheet symmetries of
compactified string theory is examined within the framework of noncommutative
geometry. The full noncommutative string spacetime is constructed using the
Fr\"ohlich-Gaw\c edzki spectral triple which incorporates the vertex operator
algebra of the string theory. The duality group appears naturally as a subgroup
of the automorphism group of the vertex operator algebra and spacetime duality
is shown to arise as the possibility of associating two independent Dirac
operators, arising from the chiral structure of the worldsheet theory, to the
noncommutative geometry.

\end{minipage}
\end{center}

\end{center}

\vskip 0.5 in

\begin{flushleft}
\end{flushleft}

\vfill
\newpage
\pagestyle{plain}
\setcounter{page}{1}

One of the key tools to the understanding of the spacetime structure
of string theory is the concept of `duality' (see \cite{Tdualrev} for
a review). A duality in string theory relates a geometry of the target
space in which the strings live to an inequivalent one. The mapping
between distinct geometries is a symmetry of the quantum string
theory. The equivalence between such geometries from the string
theoretic point of view leads to the notion of a {\it stringy} or {\it
  quantum} spacetime which forms the moduli space of string vacua and
describes the appropriate stringy modification of classical general
relativity. Dualities have been exploited recently to relate
apparently distinct string theories.
A present model for the spacetime structure of superstring theory is
`M Theory' \cite{Mdualities} which subsumes all five consistent
superstring theories in ten dimensions via duality transformations and contains
11-dimensional supergravity as its low-energy limit.

T-duality, which relates large and small radius circles, leads
to a fundamental length scale in string theory, determined by the Planck length
$l_P$. A common idea is that at distances smaller than $l_P$ the
conventional notion of a spacetime geometry is inadequate to describe its
structure. The string
configurations are conjectured to be smeared out and the notion of a `point' in
the spacetime ceases to make sense. A recent candidate theory for this picture
is the effective matrix field theory for D-branes \cite{Dbranes} in which the
spacetime coordinates are described by noncommuting matrices. In this Letter we
shall discuss how the effective string spacetime and its associated dualities
can be described using the techniques of noncommutative geometry \cite{Book}.
These mathematical tools are particularly well-suited to the study of the
structure of the stringy spacetime, in that it views it not as a set of
coordinates, but rather in terms of the set of fields defined on it.

The basic object which describes a metric space in noncommutative geometry is
the {\it spectral triple}
${\cal T}=({\cal H},{\cal A},D)$,
where $\cal H$ is a Hilbert space, ${\cal A}$ is a $C^*$-algebra of bounded
operators acting on $\cal H$, and $D$ is a Dirac operator on $\cal H$. A
spin-manifold $M$ with metric $g_{\mu\nu}$ is described by choosing
${\cal H}=L^2({\rm spin}(M))$, the square-integrable spinors on $M$, and the
{\it abelian} algebra $\alg=C^0(M)$ of continuous complex-valued functions on
$M$ acting by pointwise multiplication in $\cal H$. This is the canonical
$C^*$-algebra associated with any manifold, and it determines the topology of a
space through the continuity criterion. In fact,
there is a one-to-one correspondence between the set of all
topological spaces and the
collection of commutative $C^*$-algebras, and therefore the study of the
properties of spacetime manifolds can be substituted by a study of the
properties of abelian $C^*$-algebras. The usual Dirac operator
$D=ig^{\mu\nu}\gamma_\mu\nabla_\nu$ then describes the Riemannian geometry of
the manifold, where the real-valued gamma-matrices obey the Clifford algebra
$\left\{\gamma_\mu,\gamma_\nu\right\}=2g_{\mu\nu}$ and $\nabla_\mu$ is the
usual covariant derivative constructed from the spin-connection.
Thus, roughly speaking, $D$ is the ``inverse" of the infinitesimal $dx$
which determines geodesic distances in the spacetime.

A quantum theory of point particles on $M$ naturally supplies
${\cal H}=L^2({\rm spin}(M))$ and can therefore be thought of as describing an
ordinary spacetime $M$. In the case of string theory then, we can anticipate
that the full stringy spacetime is described by a {\it noncommutative}
geometry, {\it i.e.}\
a spectral triple with a noncommutative algebra
$\alg$. In this case the notion of a `space' breaks down, as it is in general
impossible to speak of points. In the low-energy limit of the string theory,
the vibrational modes are negligible and the
model describes ordinary point-like particles corresponding to an ordinary
(commutative) spacetime. Thus in the low-energy limit of the string
theory, ${\cal T}$ should contain a subspace
${\cal T}_0=(L^2({\rm spin}(M)),C^0(M),ig^{\mu\nu}\gamma_\mu\nabla_\nu)$
representing an ordinary spacetime manifold $M$ at large distance scales.

The symmetries of the quantum spacetime ${\cal T}$ then determine
the stringy effects on the geometry of spacetime ${\cal T}_0$
represented by ordinary quantum field theory. In the following we shall
construct the spectral triple ${\cal T}$ associated with the stringy
geometry and show that the full duality group of spacetime is determined in
a very simple fashion by the group of automorphisms of the appropriate algebra
$\alg$. The duality symmetries of the spacetime emerge as the possibility of
assigning two independent Dirac operators to the spectral triple $\cal T$
associated with strings, so that duality is simply a consequence of a
change of metric representing a redefinition of distances in the string
spacetime. These automorphisms leave the spectral triple $\cal T$ invariant and
correspond to internal symmetries of the quantum geometry. We shall also
construct the low-energy projections ${\cal T}_0$ and illustrate how
the duality isomorphisms of ${\cal T}$ correspond to the more familiar
notions of target space duality in string theory. These projections then
illuminate the full structure of the stringy modification of classical general
relativity.

The effective low-energy spectral triple has been constructed in the context of
D-brane field theory recently in \cite{Dbranesncg}. Here we shall follow the
proposal of Fr\"ohlich and Gaw\c edzki \cite{FG} that the noncommutative
geometry of the string spacetime is provided by choosing for $\alg$ the vertex
operator algebra of the underlying conformal field theory for the strings. For
illustration we shall study the simplest situation with a flat $d$-dimensional
target space of bosonic strings compactified to a torus
$T^d=(S^1)^d\cong\real^d/2\pi\Gamma$, where $\Gamma$ is a lattice of rank $d$
with metric $g_{\mu\nu}$ of Euclidean signature and with inverse $g^{\mu\nu}$.
The classical embedding functions of the strings in such a target space are the
chiral multivalued Fubini-Veneziano fields 
(in units with
$l_P=1$)
\beq
X_\pm^\mu(z_\pm)=x_\pm^\mu+ig^{\mu\nu}p_\nu^\pm\log
z_\pm+\sum_{k\neq0}\frac1{ik}\alpha_k^{(\pm)\mu}z_\pm^{-k}
\label{FVfields}\eeq
where $\mu=1,\dots,d$ and $(\alpha_k^{(\pm)\mu})^*=\alpha_{-k}^{(\pm)\mu}$.
Here we take the strings to be closed with worldsheet the
cylinder $(\tau,\sigma)\in\real\times S^1$ and $z_\pm=\e^{-i(\tau\pm\sigma)}$.
The constant modes $x_\pm^\mu$ describe the center of mass coordinates of the
string and $p_\mu^\pm$ the corresponding momenta. Singlevaluedness restricts
these left-right momenta to $p_\mu^\pm=\mbox{$\frac1{\sqrt2}$}(p_\mu\pm
d_{\mu\nu}^\pm w^\nu)$ where $\{w^\mu\}\in\Gamma$ are the winding numbers
representing the number of times that the multivalued string functions
wrap $\sigma$ around the circles of the torus $(S^1)^d$, and
$\{p_\mu\}\in\Gamma^*$ are the momenta of the winding modes in the target
space, with $\Gamma^*$ the lattice dual to $\Gamma$. The `background' matrices
$d_{\mu\nu}^\pm=g_{\mu\nu}\pm\beta_{\mu\nu}$ are constructed from the spacetime
metric $g_{\mu\nu}$ and an antisymmetric instanton tensor $\beta_{\mu\nu}$
which
does not contribute to the classical dynamics of the strings. Upon quantization
the zero modes $x_\pm^\mu,p_\mu^\pm$ become a canonically
conjugate pair and the $\alpha$'s (oscillatory modes)
act as creation and annihilation operators on a commuting pair of Fock spaces
${\cal F}^\pm$ built on vacuum states $|0\rangle_\pm$.
The Hilbert space of the string theory is thus
${\cal H}_X=L^2(T^d)^\Gamma\otimes{\cal F}^+\otimes{\cal F}^-$
where $L^2(T^d)^\Gamma=\bigoplus_{\{w^\mu\}\in\Gamma}L^2(T^d)$ are
$L^2$-spaces labelled by $w^\mu$.

We shall take the Hilbert space ${\cal H}_X$ as the first input into the
construction of the string spacetime ${\cal T}$. The next ingredient
is an algebra $\alg_X$ acting on ${\cal H}_X$ which represents its
topology. For this we exploit the operator-state correspondence which relates
the Hilbert space ${\cal H}_X$ to an appropriate operator algebra. We define
\cite{FG} $\alg_X$ to be the algebra generated by the smeared vertex operators
\beq
V_\Omega(q^+,q^-)=\int\frac{dz_+dz_-}{4\pi z_+z_-}{\cal
V}^\Omega_{q^+q^-}(z_+,z_-)
\label{smearedops}\eeq
where $q_\mu^\pm=\frac1{\sqrt2}(q_\mu\pm d_{\mu\nu}^\pm v^\nu)$, the local
vertex operators are
\bea
&{\cal V}_{q^+q^-}^\Omega(z_+,z_-)
=
:i\,V_{q^+q^-}(z_+,z_-)&\nonumber\\
&\times\prod_j\frac{r_\mu^{(+)j}}{(n_j-1)!}\partial_{z_+}
^{n_j}X_+^\mu\prod_k\frac{r_\nu^{(-)k}}{(m_k-1)!}
\partial_{z_-}^{m_k}X_-^\nu:&
\label{spinvertex}\eea
and $V_{q^+q^-}(z_+,z_-)=(-1)^{q_\mu w^\mu}:\e^{-iq^+_\mu X_+^\mu(z_+)-iq^-_\mu
X_-^\mu(z_-)}:$ are the left-right symmetric tachyon vertex operators which
form a basis for $\alg_X$. Here $\Omega=\{(r^{(+)i},r^{(-)i});n_j,m_k\}$ labels
the fields and ``:'' denotes normal ordering. The action of the operators
(\ref{smearedops}) on the unique vacuum state $|{\rm
vac}\rangle=1\otimes|0\rangle_+\otimes|0\rangle_-$ of ${\cal
H}_X$ establishes the operator-state correspondence,
implying that $\alg_X{\cal
H}_X={\cal H}_X$. For a typical state
$|\varphi_{q^+q^-}^\Omega\rangle=|q^+;q^-\rangle\otimes\prod_j
r_\mu^{(+)j}\alpha_{-n_j}^{(+)\mu}
|0\rangle_+\otimes\prod_kr_\nu^{(-)k}\alpha_{-m_k}^{(-)\nu}|0\rangle_-$ in
${\cal H}_X$, we have $|\varphi^\Omega_{q^+q^-}\rangle=V_\Omega(q^+,q^-)|{\rm
vac}\rangle$. The operator (\ref{smearedops}) is said to create a string
state of type $\Omega$ and momentum $(q^+,q^-)\in\Gamma\oplus\Gamma^*$. The
coefficients of the monomials in $z_\pm$ that arise in expansions of products
of the tachyon vertex operators
span the linear space generated by the operators (\ref{spinvertex})
as the momenta $(q^+,q^-)$ are varied. The noncommutativity
of the algebra $\alg_X$ is expressed in terms of the operator product
expansion formula of the vertex operators, which in turn encodes the complete
set of relations of the quantum field algebra and of $\alg_X$ \cite{ls}. The
vertex operators (\ref{spinvertex}) generate all string scattering amplitudes,
and so the noncommutative algebra ${\cal A}_X$ represents the ``space" of
interacting strings.

Finally, we introduce a Dirac operator. The chiral-antichiral
decomposition in (\ref{FVfields}) allows two such operators to be
introduced. The generators of target space reparametrizations
$X_\pm^\mu(z_\pm)\to X_\pm^\mu(z_\pm)+\delta X_\pm^\mu(z_\pm)$, with
$\delta X_\pm^\mu(z_\pm)$ arbitrary periodic functions, are the
conserved currents
$\delta_\pm^\mu(z_\pm)=-iz_\pm\partial_{z_\pm}X_\pm^\mu(z_\pm)$ in each chiral
sector. To define the Dirac operator, we need to augment the Hilbert space
${\cal H}_X$ to incorporate spinors. Thus we consider instead ${\cal
  H}=L^2({\rm spin}(T^d))^\Gamma\otimes{\cal F}^+\otimes{\cal F}^-$.
Locally, the spin bundle splits into two chiral parts, so that the
$L^2$-spinors split into left- and right-moving components via the
decomposition $L^2({\rm spin}(T^d))={\cal S}(T^d)^+\otimes{\cal
  S}(T^d)^-\otimes L^2(T^d)$, with ${\cal S}(T^d)^\pm$ two
anticommuting representations of the corresponding Clifford algebra with
gamma-matrices $\gamma_\mu^\pm$. The $C^*$-algebra $\alg=\id\otimes\alg_X$
acts trivially on the spinor part of $\cal H$. We then define two
mutually anticommuting Dirac operators by
\beq
D^\pm(z_\pm)=\sqrt2\gamma_\mu^\pm\delta_\pm^\mu(z_\pm)=\sum_{k=-\infty}^
\infty\sqrt2\gamma_\mu^\pm\alpha_k^{(\pm)\mu}z_\pm^{-k}
\label{diracops}\eeq
with $\alpha_0^{(\pm)\mu}=g^{\mu\nu}p_\nu^\pm$. The Dirac
operators (\ref{diracops}) are related \cite{ls} to Witten's Dirac-Ramond
operator associated with $N=1$ superstrings \cite{witten}.
These Dirac operators have been used recently by
Chamseddine in \cite{cham} to construct an effective superstring action using
the spectral action principle of noncommutative geometry \cite{specaction}.

The two natural choices of Dirac operator in the spectral triple
${\cal T}$ are thus $D=\frac1{\sqrt2}(D^++D^-)$
and $\bar D=\frac1{\sqrt2}(D^+-D^-)$. The existence of both $D$ and $\bar D$
severely restricts
the effective spacetime geometry and is intimately related to the fact that the
quantum spacetime determined by the string theory is not $T^d$,
but rather its quotient under the action of
the duality group $G$ of the string theory. The point is that there exist
several linear transformations $T:{\cal H}\to{\cal H}$ with $\bar D=TDT^{-1}$
which define automorphisms of the smeared vertex operator algebra, {\it i.e.}
$T\alg T^{-1}=\alg$. Given that the change $T$ in Dirac
operator simply corresponds to a change of metric on the spacetime, general
covariance implies that the two spacetimes represented by the
corresponding spectral triples are equivalent,
\beq
({\cal H},\alg,D)\cong({\cal H},\alg,\bar D)
\label{spectripiso}\eeq
More complicated duality symmetries can also be constructed by introducing a
larger set of Dirac operators associated with, say, an $N=2$ supersymmetric
sigma-model \cite{Tdualrev} which contains a larger symmetry than just the
chiral-antichiral one exploited above.

The spacetime duality maps $T$ are, by definition, those which preserve the
low-energy spectral triple ${\cal T}_0$. They are determined by
the various mappings between the subspaces
\beq
{\cal H}_0\equiv{\rm ker}D=\bigotimes_{\mu=1}^d\left({\cal
H}_0^{(+)\mu}\oplus{\cal H}_0^{(-)\mu}\right)
\label{spinspaces}\eeq
and those of the analogously constructed $\bar{\cal H}_0={\rm ker}\bar D$.
The kernel (\ref{spinspaces}) is composed of
$2^d$ subspaces each of which represents a particular choice of spin structure
on the $d$-torus, {\it i.e.} a choice of antiperiodic or periodic fermionic
boundary conditions around each of the $d$ circles of
$T^d$. In ${\cal H}_0^{(+)\mu}$ (resp.\ ${\cal H}_0^{(-)\mu}$), the
states satisfy
$g^{\nu\lambda}d_{\lambda\mu}^+\gamma_\nu^+=
g^{\nu\lambda}d_{\lambda\mu}^-\gamma_\nu^-$ ($\gamma_\mu^+=-\gamma_\mu^-$) and
 $p_\mu=0$ ($w^\mu=0$). On the other hand, in $\bar{\cal H}_0^{(+)\mu}$
(resp.\ $\bar{\cal H}_0^{(-)\mu}$), we have $\gamma_\mu^+=\gamma_\mu^-$
($g^{\nu\lambda}d_{\lambda\mu}^+\gamma_\nu^+=-
g^{\nu\lambda}d_{\lambda\mu}^-\gamma_\nu^-$) and $w^\mu=0$ ($p_\mu=0$). In all
of these subspaces we also have $\alpha_k^{(\pm)\nu}=0$ for all $\nu$ and all
$k>0$. This is precisely what is meant by the low-energy sector
of the string theory, the internal motion of the string is suppressed
leaving only its particle-like center of mass degrees of freedom.

We immediately find an explicit isomorphism ${\cal
H}_0^{(+)\mu}\leftrightarrow{\cal
H}_0^{(-)\mu}$ defined by $g^{\nu\lambda}d_{\lambda\mu}^\pm
\gamma_\nu^\pm\leftrightarrow\pm\gamma_\mu^\pm$
and $g^{\mu\nu}p_\nu\leftrightarrow w^\mu$,
and similarly on $\bar{\cal H}_0^{(+)\mu}\leftrightarrow\bar{\cal
H}_0^{(-)\mu}$. This means that the low-energy effective spacetime is
independent of the choice of spin structure on $(S^1)^d$.
Thus a change of spin structure
manifests itself as a T-duality symmetry of the string theory. We can therefore
make the canonical choice of antichiral low-energy subspace ${\cal
H}_0^{(-)}={\cal H}_0^{(-)1}\otimes
\cdots\otimes{\cal H}_0^{(-)d}$ determined by
$D$. In this subspace, the states carry the antichiral representation
$\gamma^+_\mu=-\gamma_\mu^-\equiv\gamma_\mu$ for all $\mu$, and they
have $w^\mu=0$ for all $\mu$ so that
$p_\mu^+=p_\mu^-=p_\mu/\sqrt2\in\Gamma^*$, as is required since in the
low-energy sector there should be no non-local string modes which wind
around the compactified directions. It is straightforward to show that
${\cal H}_0^{(-)}$ is naturally isomorphic to the Hilbert space
$L^2({\rm spin}(T^d)^-)$ of antichiral square-integrable spinors on
the torus, and that the action of the antichiral Dirac operator $D^-$
on ${\cal H}_0^{(-)}$ is $D^-|_{{\cal
    H}_0^{(-)}}=ig^{\mu\nu}\gamma_\mu\frac{\partial}{\partial x^\nu}$.

It remains to construct a low-energy projection $\alg_0$ of $\alg$.
We define $\alg_0$ to be the commutant of $D$,
\beq
\alg_0={\rm comm}D=\{V\in\alg~|~[D,V]=0\}
\label{commD}\eeq
It is the largest subalgebra of $\alg$ with the property $\alg_0{\cal
H}_0={\cal H}_0$. It can be shown that $\alg_0|_{{\cal H}_0^{(-)}}$ consists
precisely of those vertex operators $V_\Omega(q^+,q^-)$ which create string
states of identical left and right chiral momentum,
$q_\mu^+=q_\mu^-=q_\mu/\sqrt2\in\Gamma^*$. Furthermore, the smeared tachyon
generators $V(q,q)$ of $\alg_0|_{{\cal H}_0^{(-)}}$ coincide with the spacetime
functions $\e^{-iq_\mu x^\mu}$ which constitute a basis for the algebra
$C^0(T^d)$. Thus, the antichiral low-energy projection of the left-hand side of
(\ref{spectripiso}) is
\bea
{\cal T}_0^{(-)}&=&\left({\cal H}_0^{(-)},\alg_0\bigm|_{{\cal
H}_0^{(-)}},D^-\bigm|_{{\cal
H}_0^{(-)}}\right)\nonumber\\
&=&\left(L^2({\rm
spin}(T^d)^-),C^0(T^d),ig^{\mu\nu}\gamma_\mu\partial_\nu\right)
\label{T0-}\eea
which describes the toroidal spacetime $T^d$ with metric $g_{\mu\nu}$.

Now let us carry out the same construction for $\bar D$,
choosing again the corresponding antichiral low-energy subspace $\bar{\cal
H}_0^{(-)}$. In this subspace the states carry the antichiral spinor
representation
$g^{\nu\lambda}d_{\lambda\mu}^+\gamma_\nu^+=-g^{\nu\lambda}d_{\lambda\mu}^-
\gamma_\nu^-\equiv\tilde\gamma_\mu$ and have $p_\mu=0$ for all
$\mu$, so that
$(d^+)^{\mu\nu}p_\nu^+=-(d^-)^{\mu\nu}p_\nu^-=w^\mu/\sqrt2\in\Gamma$. These
conditions are T-dual to those of ${\cal H}_0^{(-)}$. In the present context,
T-duality is the linear isomorphism $T$ in (\ref{spectripiso}) that maps
$T|p^+;p^-\rangle=(-1)^{p_\mu w^\mu}|(d^+)^{-1}p^+;-(d^-)^{-1}p^-\rangle$,
$T\alpha_k^{(\pm)\mu}T^{-1}=\pm
g_{\nu\lambda}(d^\mp)^{\mu\nu}\alpha_k^{(\pm)\lambda}$, and $T\gamma_\mu^\pm
T^{-1}=g^{\nu\lambda}d_{\mu\nu}^\mp\gamma_\lambda^\pm$. This
implies that the target space metric changes to its dual
$\tilde g^{\mu\nu}=(d^+)^{\mu\lambda}g_{\lambda\rho}(d^-)^{\rho\nu}$ which
defines a metric on $\Gamma^*$. $T$ acts on the smeared vertex
operator algebra as $TV_\Omega(q^+,q^-)T^{-1}=(-1)^{q_\mu
w^\mu}V_\Omega(q^+(d^+)^{-1},-q^-(d^-)^{-1})$, and so this linear mapping
preserves both $\cal H$ and $\alg$ at the same time as establishing
the spectrum-preserving mapping $\bar D=TDT^{-1}$. This transformation is the
noncommutative geometry version of the celebrated T-duality transformation of
string theory which exchanges the torus $T^d$ with its dual
$(T^d)^*\cong\real^d/2\pi\Gamma^*$, and interchanges momenta
and windings in the spectrum.
It corresponds to an inversion of the background matrices
$d^\pm\to(d^\pm)^{-1}$ and is the $d$-dimensional analog of the $R\to1/R$
circle duality \cite{Tdualrev}.

In the low-energy sector, it is once again possible to show that $\bar{\cal
H}_0^{(-)}=L^2({\rm spin}^*(T^d)^-)$, that $D^-|_{\bar{\cal H}_0^{(-)}}=i\tilde
g^{\mu\nu}\tilde\gamma_\mu\frac\partial{\partial x^\nu}$ and
$\bar\alg_0|_{\bar{\cal H}_0^{(-)}}={\rm comm}\bar D|_{\bar{\cal
H}_0^{(-)}}=C^0(T^d)$, yielding the T-dual low-energy triple
$\bar{\cal T}_0^{(-)}$ to (\ref{T0-}). Then according to (\ref{spectripiso}),
as subspaces of the effective quantum spacetime we have the
equivalence between $(L^2({\rm
spin}(T^d)^-),C^{\infty}(T^d),ig^{\mu\nu}\gamma_\mu\partial_\nu)$ and
$(L^2({\rm spin}^*(T^d)^-),C^{\infty}(T^d),i\tilde
g^{\mu\nu}\tilde\gamma_\mu\partial_\nu)$ which is the usual statement of the
T-duality symmetry
$T^d\leftrightarrow(T^d)^*$ of bosonic string theory compactified on a
$d$-torus. Notice how the statement that this duality symmetry corresponds to
the target space symmetry $g_{\mu\nu}\leftrightarrow\tilde g^{\mu\nu}$ and the
symmetry under interchange of momentum and winding in the compactified
string spectrum arise very naturally from the point of view of noncommutative
geometry. The crucial aspect of the low-energy projections is that the
restricted vertex operator subalgebras consist of only the zero mode components
of the tachyon vertex operators in (\ref{spinvertex}).

There are many other symmetries that arise from such isomorphisms,
corresponding to mappings from ${\cal H}_0^{(-)}$ onto other subspaces of ${\rm
ker}\bar D$. For example, worldsheet parity and factorized duality
(equivalently mirror symmetry when $d$ is even) easily arise in this way
\cite{ls}.
Besides these geometrical symmetries, there are two other internal symmetries
of the spacetime which trivially leave the spectral triples in
(\ref{spectripiso}) invariant. The first is a change of basis of the lattice
$\Gamma$, a process which preserves all quantities composing the spectral
triples. The second is the global shift
$\beta_{\mu\nu}\to\beta_{\mu\nu}+C_{\mu\nu}$ of the instanton tensor by an
antisymmetric integer-valued matrix $C_{\mu\nu}$, which can be absorbed in a
shift of the momenta $p_\mu\to p_\mu-C_{\mu\nu}w^\nu$ and
leaves the spectral triples unaffected. It can be shown that the set of duality
transformations described here generate the discrete duality group of the
string theory which is the semidirect product $G=O(d,d;\zed)\otimes_{\rm
S}\zed_2$ of the lattice automorphism group $O(d,d;\zed)$ and the reflection
group $\zed_2$ corresponding to worldsheet parity. The above constructions thus
show how the geometry and topology relevant for general relativity must be
embedded in a larger noncommutative structure in string theory. They can also
be generalized to include worldsheet gravitational effects and more complicated
target space geometries \cite{ls}, as well as to obtain duality maps between
different spectral triples representing the equivalences of distinct string
theories and
D-brane models as dictated by M Theory \cite{Mdualities}. The duality group is
a subgroup of the full group of automorphisms of the algebra $\alg$ which
represents the diffeomorphism and internal symmetries of the stringy general
relativity. In fact, $G$ appears in terms of the zero modes of the Dirac
operators from (\ref{spinspaces}), while the full set of automorphisms of the
spectral triple $\cal T$ are determined by mappings between different Dirac
operators that have the same spectrum \cite{specaction}. Thus the full set of
symmetries of the string spacetime is intimately connected with some of the
exotic mathematical structures which appear in the theory of vertex operator
algebras, such as the Monster Lie algebra \cite{vertex}. More details about
these constructions will appear in a forthcoming paper \cite{ls}.

\end{document}